\documentclass[prl,superscriptaddress,twocolumn,amsmath,amssymb,nofootinbib,longbibliography]{revtex4-2}

\usepackage{graphicx}
\usepackage{amsmath}
\usepackage{physics}
\usepackage{color}
\usepackage{slashed}
\usepackage{tensor}
\usepackage[utf8]{inputenc}
\usepackage[normalem]{ulem}

\newcommand{\ba}{\begin{eqnarray}}
\newcommand{\ea}{\end{eqnarray}}

\newcommand{\be}{\begin{equation}}             
\newcommand{\ee}{\end{equation}}               
\newcommand{\notprop}{\propto\kern-1\@ptsize pt \diagup}

\begin{document}

\title{Thermodynamics of AdS Black Holes: Central Charge Criticality}

\author{Wan Cong}
\email{wcong@uwaterloo.ca}	
	\affiliation{Department of Physics and Astronomy, University of Waterloo,
		Waterloo, Ontario, Canada, N2L 3G1}	
	\affiliation{Perimeter Institute, 31 Caroline Street North, Waterloo, ON, N2L 2Y5, Canada}

\author{David Kubiz\v n\'ak}
\email{dkubiznak@perimeterinstitute.ca}
	\affiliation{Perimeter Institute, 31 Caroline Street North, Waterloo, ON, N2L 2Y5, Canada}
	\affiliation{Department of Physics and Astronomy, University of Waterloo,
		Waterloo, Ontario, Canada, N2L 3G1}

\author{Robert B. Mann}
	\email{rbmann@uwaterloo.ca}
		\affiliation{Department of Physics and Astronomy, University of Waterloo,
		Waterloo, Ontario, Canada, N2L 3G1}	
	\affiliation{Perimeter Institute, 31 Caroline Street North, Waterloo, ON, N2L 2Y5, Canada}

\date{May 5, 2021}
	
	\begin{abstract}
		We reconsider the thermodynamics of  AdS black holes in the context of gauge--gravity duality. In this  new setting where both the cosmological constant $\Lambda$ and the gravitational Newton constant $G$  are varied in the bulk, we rewrite the first law in a new form   containing both $\Lambda$ (associated with thermodynamic pressure) and the central charge $C$ of the dual CFT theory and their conjugate variables. We obtain a novel thermodynamic volume, in turn leading to a new understanding of the Van der Waals behavior of the charged AdS black holes, in which phase changes are governed by the degrees of freedom in the CFT.  Compared to the `old' $P-V$ criticality, this new criticality is `universal' (independent of the bulk pressure) and directly relates to the thermodynamics of the dual field theory and its central charge.    
	\end{abstract}

\maketitle

Black holes and their thermodynamics  have been of crucial importance in providing clues about the nature of quantum 
gravity. Of particular interest are asymptotically anti de Sitter (AdS) black holes, which provide a description of the dual conformal field theory (CFT) at finite temperature via the AdS/CFT correspondence \cite{Maldacena:1997re}. Such black holes can be in thermal equilibrium with their Hawking radiation and exhibit interesting thermodynamic phase transitions, such as the first order Hawking--Page phase transition \cite{Hawking:1982dh} which corresponds to the confinement/deconfinement of the dual quark gluon plasma \cite{Witten:1998zw}, or the existence of a second order Van der Waals type phase transition  for charged AdS black holes  \cite{Chamblin:1999tk, Chamblin:1999hg, Cvetic:1999ne,Kubiznak:2012wp}.

One of the more interesting developments  in recent years has been a realization that a negative cosmological constant $\Lambda$ induces a positive thermodynamic pressure $P$,
\be\label{P}
P=-\frac{\Lambda}{8\pi G}\,,\quad \Lambda=-\frac{(D-1)(D-2)}{2 l^2}\,, 
\ee
upon which the thermodynamics of AdS black holes becomes more complete, `identical' to thermodynamics of ordinary systems (including the $P\delta V$ term) \cite{Kubiznak:2012wp}. 
Here, $l$ is the radius of the $D$-dimensional AdS space and $G$ is the (dimensionful) Newton  gravitational constant;
 we set $\hbar=c=1$.

In this {\em extended thermodynamic phase space} \cite{Kastor:2009wy},  the black hole mass $M$ is  interpreted as thermodynamic enthalpy rather than internal energy, and the  first law of thermodynamics and the corresponding Smarr relation for a black hole of charge $Q$, surface gravity $\kappa$, angular momentum $J$, and area $A$ are
\ba
\delta M&=&T\delta S+V\delta P+\phi \delta Q+\Omega \delta J\,,\label{flaw}\\
M&=&\frac{D-2}{D-3}(TS+\Omega J)+\phi Q-\frac{2}{D-3}PV\,,\label{smarr}
\ea 
where entropy $S$ and temperature $T$  are
\be
S=\frac{A}{4G}\,,\quad T=\frac{\kappa}{2\pi}\,, 
\ee
with $\phi,\Omega$ the respective conjugates to $Q,J$, and  where
\be\label{V}
V=\Bigl(\frac{\partial M}{\partial P}\Bigr)_{S,Q,J}\,
\ee
is the thermodynamic volume  conjugate to $P$ \cite{Dolan:2010ha, Cvetic:2010jb}.

In this framework,  black hole thermodynamics is phenomenologically much richer than previously expected,  with black holes exhibiting Van der Waals \cite{Kubiznak:2012wp}, re-entrant \cite{Altamirano:2013ane}, superfluid \cite{Hennigar:2016xwd}, and polymer-type phase transitions \cite{Dolan:2014vba}, along with triple points \cite{Altamirano:2013uqa} and universal scaling behaviour of the Ruppeiner curvature \cite{Wei:2019uqg}.  For these reasons this subdiscipline has come to be called {\em Black Hole Chemistry} \cite{Kubiznak:2016qmn}.

The {\em holographic interpretation} of black hole chemistry has been somewhat  elusive  \cite{Johnson:2014yja, Dolan:2014cja, Kastor:2014dra, Zhang:2014uoa, Zhang:2015ova, Dolan:2016jjc, McCarthy:2017amh}.  The first law \eqref{flaw} cannot be straightforwardly related to the corresponding thermodynamics of the holographic dual field theory \cite{Karch:2015rpa, 
Sinamuli:2017rhp, Visser:2021eqk} because variations of the bulk cosmological constant $\Lambda$ correspond to changing both  the central charge $C$ (or the number of colours $N$) and   the CFT volume ${\cal V}$.  Indeed, it furthermore corresponds to changing the notion of electric charge and the corresponding chemical potential that both rescale with the AdS radius $l$. It is possible to hold $C$ fixed so that the field theory remains the same (in which case  varying $\Lambda$ has the more natural consequence of varying ${\cal V}$ of the CFT) by simultaneously varying Newton’s constant $G$. (Alternative reasons for as to why $G$ should be varied in the first law of black hole thermodynamics have already been put forward in \cite{Kastor:2010gq}.)

Here we demonstrate that variation of $G$ has profound consequences for   black hole chemistry and its holographic interpretation. We build on previous holographic generalizations   
\cite{Visser:2021eqk}  of the first law \eqref{flaw} which include variations of $G$
by rewriting it in a new (mixed) form -- in terms of variations of the bulk pressure $P$ and the CFT central charge $C$. 
This allows for a study of bulk thermodynamics yielding a new definition for the thermodynamic black hole volume whilst it is possible to  restrict to the same CFT (fixed $C$) on the AdS boundary. 
Surprisingly, as we shall see this also leads to a novel understanding of the phase transitions of AdS black holes which, contrary to the conclusions of black hole chemistry, no longer necessarily depends on the bulk critical pressure but rather derives from the critical central charge of the dual CFT.

We begin by considering the correspondence between the holographic CFT and bulk thermodynamic quantities valid in Einstein gravity (large $N$ limit) \cite{Karch:2015rpa, Visser:2021eqk} .  For the CFT the first law is \cite{Visser:2021eqk}
\be\label{FirstHol}
\delta E=T\delta S-p d{\cal V}+\tilde\phi \delta \tilde Q+\Omega \delta J+ \tilde \mu \delta C \,, 
\ee
where $E$ is the CFT energy (not entalphy),  $p$ and ${\cal V}={\cal V}_0 l^{D-2}$ are the CFT pressure and volume,  $\tilde \mu$ is the chemical potential for the central charge $C$, which is proportional to $N$ to some power ($C\propto N^2$ for $SU(N)$ gauge theories with conformal symmetry), $J$ and $\Omega$ are the angular momentum and conjugate angular velocity, and $ \tilde Q, \tilde\phi$ are its respective holographic charge and conjugate potential. These quantities scale as
\ba
[E]&=&[T]=[\Omega]=[\tilde \mu]=\frac{1}{L}\,,\quad [{\cal V}]=L^{D-2}\,,\nonumber\\
{}[S]&=&[\tilde Q]=[J]=[C]=L^0\,.
\label{units1}
\ea
which in turn implies
\be
E = (D-2)  p {\cal V}  
\ee
from  the standard   dimensional Euler scaling argument.  

Using the fact that $E,S,\tilde Q,J, C$ scale as $C$, but ${\cal V}$ does not \cite{Visser:2021eqk}, 
a similar Euler argument yields
\be\label{SmarrHol}
E=TS+\tilde \phi \tilde Q+\Omega J+\tilde\mu C\,, 
\ee
 which is  the {\em holographic Smarr}   relation  \cite{Visser:2021eqk}.
Note that  \eqref{SmarrHol} has  no $D$-dependent factors, and that  the $p-{\cal V}$ term does not appear here.

In order to obtain a duality between holographic and bulk thermodynamics, we employ the following duality relation:
\be\label{C}
C=k \frac{l^{D-2}}{16\pi G}\,, 
\ee
where the numerical factor $k$ depends on the details of the particular holographic system  \cite{Karch:2015rpa}. 
We likewise identify 
\be\label{Qt}
E=M\,,  \qquad \tilde Q= \frac{Ql}{\sqrt{G}}\,,\qquad \tilde \phi = \frac{\phi\sqrt{G}}{l}\,,
\ee
and make use of \eqref{P} to identify $\Lambda$ and $l$.  
Allowing both $l$ and $G$ to vary \cite{Visser:2021eqk}, from   \eqref{FirstHol} and \eqref{SmarrHol} we obtain
\begin{align}
\delta (GM) & =\frac{\kappa}{8\pi } \delta A+\Omega\; \delta( GJ)+ \sqrt{G} {\phi}\; \delta( \sqrt{G} Q) -\frac{V}{8\pi} \delta \Lambda  \nonumber\\
 \Leftrightarrow \quad \delta M & =\frac{\kappa}{8\pi G} \delta A+\Omega \delta J+\phi \delta Q -\frac{V}{8\pi G} \delta \Lambda-\alpha \frac{\delta G}{G}\,, 
 \label{Bulk1}
\end{align}
where
\ba 
\alpha&=&\frac{1}{2}\phi Q+\tilde \mu C+TS=M-\Omega J-\frac{1}{2}\phi Q\,,
\label{alp}
\\
V&=&\frac{8\pi G l^2}{(D-1)(D-2)}\Bigl(M-\phi Q-(D-2)C\tilde \mu\Bigr)\,\nonumber\\
&=&\frac{(D-3)}{2P}\Bigl(\frac{D-2}{D-3}(\Omega J+TS)+\phi Q-M\Bigr)\,.
\label{Vsmarr}
\ea
  We see that when $G$ is held fixed the first law \eqref{Bulk1} reduces to \eqref{flaw}. 
 
 The quantity
 $\alpha/G$ is  conjugate to $G$, and the factor  in front of the  $\phi Q$ term in \eqref{alp} differs from that presented previously     \cite{Visser:2021eqk} due to the $G$-dependence of $\tilde Q$ and $\tilde \phi$ of the dual CFT, c.f. \eqref{Qt}.
 Furthermore,  \eqref{Vsmarr} is equivalent to the bulk Smarr relation \eqref{smarr}. 
  
Since $G$ is now a thermodynamic variable, dimensional analysis implies 
\begin{align}\label{units2} 
&[\phi] =  L^{\frac{2-D}{2}}\,, \quad [Q]=L^\frac{D-4}{2} \,,
\quad [A]=[G]=L^{D-2}\,, 
\\{ }
& [M]= \frac{1}{L}\,,\quad  
[\Lambda] = \frac{1}{L^2}\,, \quad [P]=\frac{1}{L^D}\,, \quad
[V]= L^{D-1}\,,  \nonumber
\end{align} 
from which one can recover the bulk Smarr relation \eqref{smarr} provided $\alpha$ is given as in \eqref{alp}. Notice that this way of recovering \eqref{smarr} required input from the CFT side since \eqref{alp} follows from the holographic Smarr relation. An alternative way without going through the CFT would be to make use of the dependence
\be
\label{gdep}
G M = {\cal M}(A,\sqrt{G}Q,GJ,\Lambda)\,
\ee
in the bulk. Its differential, together with the first law \eqref{Bulk1} then yields the expression for $\alpha$, \eqref{alp}, which in turn yields \eqref{smarr} by the usual dimensional scaling argument.

We would now like to rewrite \eqref{Bulk1} in terms of variations in the bulk pressure $P$. We can do so by making use of 
\be
\label{deltaG}
\frac{\delta G}{G}=-\frac{2}{D}\frac{\delta C}{C}-\frac{D-2}{D}\frac{\delta P}{P}\,,
\ee 
which follows from \eqref{C}. Substituting this into \eqref{Bulk1} allows us to write the bulk first law in  the following `mixed' form: 
\be
\delta M = T \delta S+\Omega \delta J+\phi \delta Q+V_C \delta P+\mu \delta C\,, \label{FirstC} 
\ee
where
\be\label{VC}
V_C =\frac{2M+(D-4)\phi Q}{2DP}\,,\quad  \mu =\frac{2P(V_C-V)}{C(D-2)}\,,
\ee
are  the new thermodynamic volume $V_C$ and chemical potential  $\mu$. Note from \eqref{deltaG} that fixing the central charge determines variations of $G$ in terms of those in $P$, whereas fixing $G$ determines variations of $C$ in terms of those in $P$ (while the CFT volume ${\cal V}$ changes as well) which then recovers the usual thermodynamics in black hole chemistry \eqref{flaw}.
 
It is well known that for a large class of black holes, the thermodynamic volume $V$, \eqref{V}, satisfies the reverse isoperimetric inequality \cite{Cvetic:2010jb} 
\be\label{ISO}
{\cal R} \equiv \Bigl(\frac{(D-1)V}{\omega_{D-2}}\Bigr)^{\frac{1}{D-1}}\Bigl(\frac{\omega_{D-2}}{A}\Bigr)^{\frac{1}{D-2}} \geq 1\,, 
\ee
where $A$ is the horizon area and $\omega_n={2\pi^{\frac{n+1}{2}}}/{\Gamma\bigl(\frac{n+1}{2}\bigr)}$,
with the inequality saturated for (simple) spherical black holes. For fixed $G$ this amounts to a statement that  
 Schwarzschild AdS black holes are ‘maximally entropic’ for a given black hole volume $V$. It follows from \eqref{VC} 
(for spherical black holes) that whenever
the chemical potential $\mu$ becomes negative, that is $V>V_C$, the standard (rather than reverse) isoperimetric inequality for $V_C$ is satisfied. As we shall see below, this happens for large charged AdS black holes -- with respect to the volume $V_C$, such black holes are therefore `superentropic' \cite{Hennigar:2014cfa}.


We now test our new notions on a charged AdS black hole, reconsidering its thermodynamic criticality.
First studied   in the $\phi-Q$ plane \cite{Chamblin:1999tk, Chamblin:1999hg, Cvetic:1999ne},  the analogy between the thermodynamics of charged AdS black holes and Van der Waals fluids was properly completed in the extended phase space \cite{Kubiznak:2012wp}. Subsequently the corresponding CFT criticality in the context of $AdS_5\times S_5$  was studied  
\cite{Dolan:2014cja, Zhang:2014uoa, Zhang:2015ova} (see also \cite{Chabab:2015ytz}) in the $\tilde \mu-C$ plane and also in the $p-{\cal V}$ plane \cite{Dolan:2016jjc}. Maxwell's equal area law for the entanglement entropy has likewise been investigated \cite{McCarthy:2017amh}.

A charged AdS black hole is characterized by the following 
metric and gauge potential
\ba
ds^2&=&-fdt^2+\frac{dr^2}{f}+r^2 d\Omega_2^2\,, \quad
A = -\frac{Q}{r}dt\,, \nonumber\\
f&=&1-\frac{2GM}{r}+\frac{G Q^2 }{r^2}+\frac{r^2}{l^2}\,, 
\ea
from which we find 
\ba\label{TDs}
M&=&\frac{r_+(l^2+r_+^2)}{2l^2 G}+\frac{Q^2}{2r_+}\,,\quad T=\frac{3r_+^4+l^2r_+^2-GQ^2l^2}{4\pi l^2r_+^3}\,,\nonumber\\
S&=&\frac{\pi r_+^2}{G}\,, \quad V=\frac{4\pi r_+^3}{3}\,,\quad 
\phi=\frac{Q}{r_+}\,,
\ea
for the standard thermodynamic quantities, while the new thermodynamic variables are
\ba
V_C&=&\frac{\pi}{3 r_+}\Bigl(GQ^2l^2+l^2r_+^2+r_+^4\Bigr)\,, \\
\mu&=&
\frac{2\pi}{k l^4r_+}\Bigl(GQ^2l^2+l^2r_+^2-3r_+^4\Bigr)\,. 
\ea

\begin{figure}
\begin{center}
\rotatebox{0}{
\includegraphics[width=0.49\textwidth,height=0.35\textheight]
{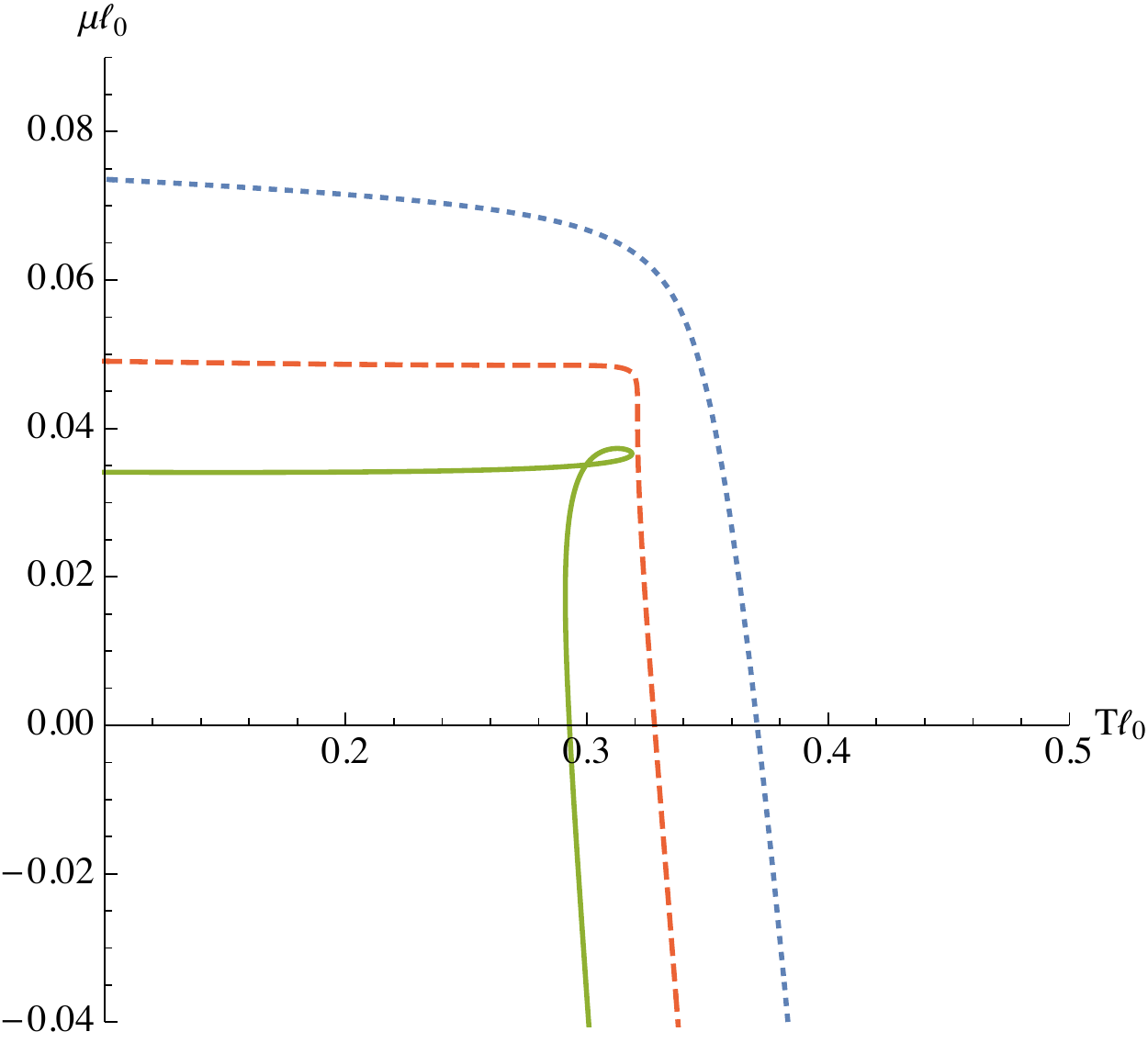}
}
\caption{{\bf $\mu-T$ diagram.} The chemical potential is displayed as a function of $Tl_0$ for fixed $P \ell_0^4=10$ and $C=20$ (blue dotted curve), $C=C_c=36$ (red dashed curve) and $C=60$ (green solid curve). Here, $Q=1, k=16\pi$,  and $\ell_0$ is an arbitrary reference length scale.  
}  \label{fig:FigmuTQ}
\end{center}
\end{figure}

Small black holes, for which
\be
r^2_+ < \frac{l^2}{6}\left(1+\sqrt{1+\frac{12 GQ^2}{l^2}}\right)\,, 
\ee
have positive $\mu$ (typical for quantum regime).
 Large black holes have large negative $\mu$ (classical regime) and violate the reverse isoperimetric inequality for  
$V_C$ (while this is saturated for $V$). We display the corresponding  $\mu-T$ diagram (for fixed $P$ and $Q$) and various values of $C$ in Fig.~\ref{fig:FigmuTQ}. The multi-valuedness for $C>C_c$ indicates the presence of the first order phase transition (see below).

\begin{figure}
\begin{center}
\rotatebox{0}{
\includegraphics[width=0.49\textwidth,height=0.35\textheight]
{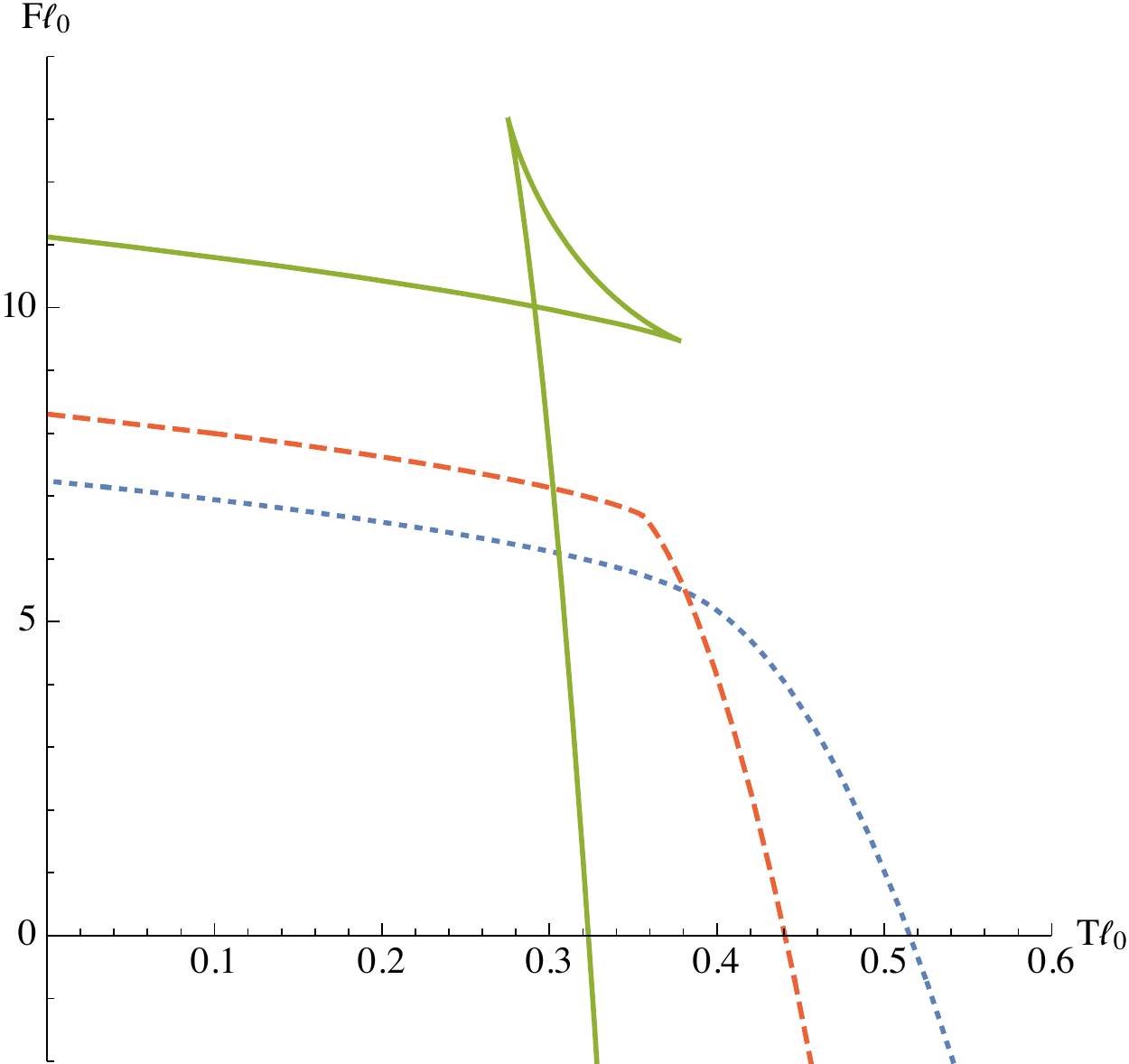}
}
\caption{{\bf Free energy diagram.} $F-T$ diagram is displayed as a function of $C$ for fixed $P=15 \ell_0^4$. The dashed red corresponds to $C=C_{crit}=36$, the dotted blue is for $C=20$ and the solid green for $C=120$. Only theories with large number of freedom admit a first order phase transition. The temperature of the critical point depends on $P$. We have set $Q=1$ and $k=16\pi$.
}  \label{fig:FigFEn}
\end{center}
\end{figure}

To investigate critical behavior, we study the free energy $F=F(T, P, Q, C)$, given by 
\begin{align} \label{Fcharged} 
F&=M-TS=\frac{3GQ^2l^2+l^2r_+^2-r_+^4}{4Gr_+l^2}  \nonumber\\
&=   \frac{2\pi r_+}{3}\sqrt{\frac{6PC}{k}}-\frac{2\pi P r_+^3}{3}+\frac{3Q^2}{4r_+}\,,
\end{align}
 where we have employed relations \eqref{P} and \eqref{C} and $r_+=r_+(T,P,Q,C)$ is implicitly given by the second relation \eqref{TDs}.
Fixing $Q$, the corresponding $F=F(T, P, Q, C)$ can be plotted parametrically, with dimensionful quantities displayed in units of some fiducial length $\ell_0$. What matters for criticality is, in contrast to   standard  black hole chemistry \cite{Kubiznak:2016qmn}, 
 the value of the central charge -- there exists a critical central charge $C_{c}$, above which ($C>C_{c}$) we observe  swallowtail behaviour (for any $P$),  whereas for $C<C_c$ the free energy is smooth (for any $P$).  In other words,  there is a first order phase transition for theories with large central charge (or many degrees of freedom), but no such phase transition for theories with small central charge (or few degrees of freedom), see Fig.~\ref{fig:FigFEn}.

\ To find this critical central charge, we impose  
\be\label{Frr}
\frac{\partial T}{\partial r_+}=0=\frac{\partial^2 T}{\partial r_+^2}\,, 
\ee 
corresponding to the ``meeting'' of the two cusps on the $F-T$ diagram. This  yields 
\be\label{Cc}
C_c=\frac{9k Q^2}{4\pi}\,, 
\ee 
for the critical central charge,
which is `universal' -- completely determined from $Q$ and independent of the thermodynamic pressure $P$.

In other words, charged AdS black holes demonstrate $\mu-C$ criticality with the corresponding equation of state
\be
C=C({\mu}, T, P, Q)\,,
\ee
given implicitly by \eqref{P}, \eqref{C},  and \eqref{TDs}. In the vicinity of the critical point, the phase transition in the $\mu-C$ plane follows the corresponding Maxwell's equal area law \cite{inprep}.

\begin{figure}
\begin{center}
\rotatebox{0}{
\includegraphics[width=0.49\textwidth,height=0.35\textheight]
{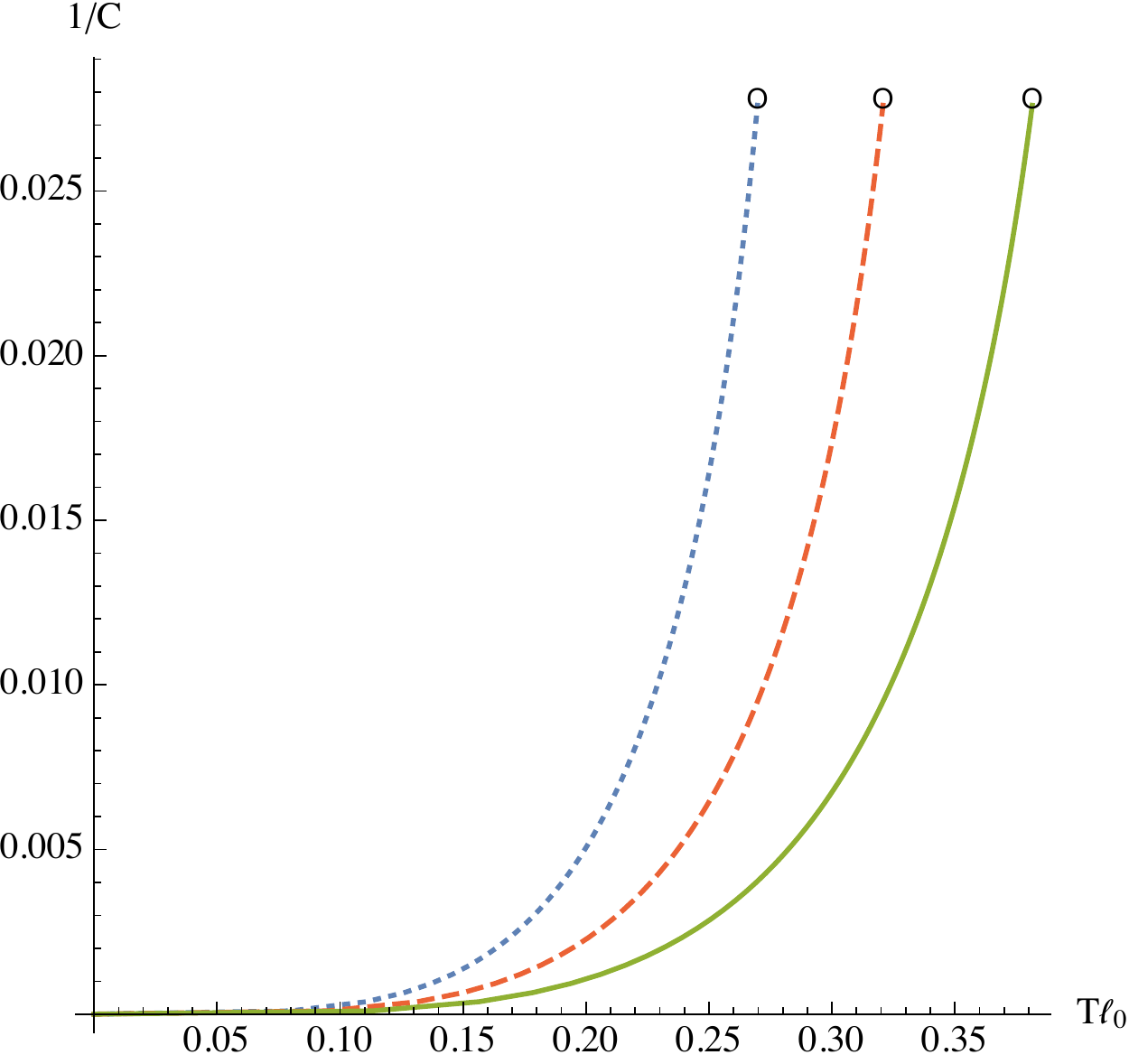}
}
\caption{{\bf Phase diagram.} The coexistence curves in the $1/C-T$ phase diagram are displayed for $P\, \ell_0^4=20$ (solid green) $P\, \ell_0^4=5$ (dotted blue) and $P\, \ell_0^4=10$ (dashed red). These curves terminate at the corresponding critical points which all share the same critical $C_c=36$ while the corresponding $T_c$ vary with pressure.  
}  
\label{fig:CT}
\end{center}
\end{figure}

 While $C_c$ is universal, other characteristics of the critical point depend on the pressure. Namely, \eqref{Frr} yields  
\be\label{rc}
r_c=\sqrt{6G} Q\,,\quad l_c=6 \sqrt{G} Q\,, \quad 
T_c=\frac{\sqrt{6}}{18 \sqrt{G} \pi Q}\,,
\ee
which (apart from explicit factors of $G$) coincides
with critical quantities obtained previously \cite{Kubiznak:2012wp}. 
However, now we have a family of critical points as $G$ can be independently varied. Using alternatively \eqref{P} we obtain 
\be\label{Gc}
G=\frac{1}{\sqrt{96 \pi P}Q}\,,
\ee   
and so we can vary $P$ to obtain this family.
Other critical quantities can easily be expressed using \eqref{rc} and \eqref{Gc}.
In Fig.~\ref{fig:CT} we plot the coexistence curves in the $1/C$--$T$ plane.   We see that the corresponding critical points all share the same critical value of $C$, whereas the corresponding $T_c$ varies with pressure.


Summarizing, we have demonstrated that gauge-gravity duality provides a new understanding of the phase behaviour of charged  AdS black holes.  We find that 
phase behaviour is governed by a critical value of the central charge $C$ instead of the pressure. For a  boundary theory with  many degrees of freedom (large central charge) there is a bulk first order phase transition, whereas for one with few degrees of freedom (small central charge) there is not. 
The universality of the central charge (its independence of the bulk pressure) can be expected from dimensional analysis. It follows that the critical behavior of rotating black holes in any number of dimensions should also be universal. On the other hand, however, we may expect that for charged black holes in dimensions other than four, the critical central charge will depend on the bulk pressure.

We stress that there are important distinctions between the  bulk $\mu-C$ criticality we have studied compared to previous investigations of holographic CFT \cite{Dolan:2014cja, Zhang:2014uoa, Zhang:2015ova, Chabab:2015ytz}, which employed the particular setting of $AdS_5\times S_5$ (or $AdS_4\times S_7$), and which  derived $G$  from the fundamental (fixed) higher-dimensional theory by volume compactification. This leaves only one independent variation of $G$ and $l$. Futhermore, the CFT volume cannot vary (as the corresponding densities were used) and the CFT first law differs from  \eqref{FirstHol}. 
Our approach, based on \eqref{C}, is more general:  $G$ and $l$ vary independently.

We conclude with a number of open questions. For example, in this letter we have concentrated on the mixed bulk first law \eqref{FirstC} where $P$ and $C$ were varied. What if instead one considered the variations of $G$ and $C$? In either setting, what happens to the more interesting 
phase transitions discovered in the black hole chemistry such as the re-entrant phase transitions or the isolated critical points? Are they still present? And if they are, are they governed by the universal critical central charge? What happens when the duality relation \eqref{C} has to be generalized? Can we also incorporate the $1/N$ corrections? These and many other questions remain interesting subjects for future investigation.

\section*{Acknowledgements}
\label{sc:acknowledgements}

This work was also supported by the Perimeter Institute for Theoretical Physics and by the Natural Sciences and Engineering Research Council of Canada (NSERC). Research at Perimeter Institute is supported in part by the Government of Canada through the Department of Innovation, Science and Economic Development Canada and by the Province of Ontario through the Ministry of Colleges and Universities. 
Perimeter Institute and the University of Waterloo are situated on the Haldimand Tract, land that was promised to the Haudenosaunee of the Six Nations of the Grand River, and is within the territory of the Neutral, Anishnawbe, and Haudenosaunee peoples.


\providecommand{\href}[2]{#2}\begingroup\raggedright\endgroup

\end{document}